\documentclass{svmult}

\usepackage{makeidx}     
\usepackage{graphicx}    
\usepackage{multicol}    

\def\hMpc{\ifmmode{h^{-1}{\rm Mpc}}\else{$h^{-1}{\rm Mpc}$}\fi}
\def\hkpc{\ifmmode{h^{-1}{\rm kpc}}\else{$h^{-1}{\rm kpc}$}\fi}
\def\hMsun{\ifmmode{h^{-1}M_\odot}\else{$h^{-1}M_\odot$}\fi}

\begin{document}

\title*{Simulations of the Local Universe}
\author{S. Gottl\"ober\inst{1}\and
A. Klypin \inst{2}\and  A. Kravtsov\inst{3} \and Y. Hoffman\inst{4} 
\and A. Faltenbacher\inst{1}}
\institute {Astrophysikalisches Institut Potsdam, An der Sternwarte 16, 14482
Potsdam, Germany
\textit{sgottloeber@aip.de}
\and Astronomy Department, New Mexico State University, USA
\textit{aklypin@nmsu.edu}
\and Dept. of Astronomy \& Astrophysics, CfCP,
The University of Chicago, USA
\and Racah Institute of Physics, Hebrew University, Jerusalem, Israel 
}
%
%
\maketitle

\smallskip
\section{Introduction}
\label{sec:1}  

One of the greatest challenges of modern astrophysics is understanding
how galaxies,  such as our Milky  Way, form within the  framework of the
Big  Bang cosmology. The  current theory  of structure  formation, the
extension  of the Big  Bang model  called the  Cold Dark  Matter (CDM)
scenario,  predicts  that galaxies  form within extended
massive dark matter halos built from  smaller
pieces that  collided and  merged, resulting in the hierarchy of 
galaxies, groups, and clusters observed today.  The
entire  sequence  of  events  is  thought  to  be  seeded  by  quantum
fluctuations  in the very  early Universe  and governed  by mysterious
"dark  matter"  which constitutes  about  85\%  of  all matter  in  the
universe.  Although the accurate properties of galaxies depend on complicated
baryonic processes (radiative cooling, formation and evolution of stars, etc.)
operating on small scales, we expect that overall spatial 
distribution of dark matter halos is closely related to the observed galaxy
distribution. Here we present numerical simulations designed to
study the formation, evolution and present day properties of such dark matter
halos in different cosmological environments.

In all simulations the spatially flat cold dark matter model with a
cosmological constant ($\Lambda$CDM with $\Omega_{\rm M}=0.3$,
$\Omega_{\Lambda}=0.7$, $\sigma_8=0.9$, and $h=0.7$), favored
by most current observations, has been assumed.

\smallskip

\section{Numerical simulations}
\label{sec:2}

The Adaptive Refinement Tree (ART) $N$-body code
\cite{kkk97,kravtsov99} was used to run all numerical simulation
analyzed in this paper. In some of the simulations described below the
code also included eulerian gasdynamics \cite{kkh02}. The code uses 
Adaptive Mesh Refinement technique to achieve high resolution in the
regions of interests. The computational box is covered with a uniform
grid which defines the lowest (zeroth) level of resolution.  
The code then reaches high force resolution by recursively refining all
high density regions using an automated refinement algorithm.  
This creates an hierarchy of refinement meshes of different
resolutions, sizes, and geometries covering regions of interest. The 
refinement data structures and algorithms \cite{khokhlov98}
allow individual cubic cells to be refined. The shape of the refinement
mesh can thus effectively match the geometry of the region
of interest. This algorithm is well suited for simulations of a
selected region within a large computational box, as in the
simulations presented below.

During the integration, spatial refinement is accompanied by temporal
refinement.  Namely, each level of refinement, $l$, is integrated with
its own time step $\Delta a_l=\Delta a_0/2^l$, where $\Delta a_0$ is
the global time step of the zeroth refinement level.  This variable
time stepping is very important for accuracy of the results.  As the
force resolution increases, more steps are needed to integrate the
trajectories accurately. In addition to spatial and temporal
refinement, simulations described below also use non-adaptive mass
refinement to increase the mass (and correspondingly the force)
resolution inside a specific region.  The multiple mass resolution is
implemented in the following way \cite{kkbp01}.  We first set up a
realization of the initial spectrum of perturbations in such a way
that initial conditions for a large number ($1024^3$) of particles can
be generated in the simulation box.  Initial coordinates and
velocities of the particles are then calculated using all waves
ranging from the fundamental mode $k=2\pi/L$ to the Nyquist frequency
$k=2\pi/L\times N^{1/3}/2$, where $L$ is the box size and $N$ is the
number of particles in the simulation.  Particles outside
high-resolution regions are then merged into particles of larger mass
and this process can be repeated for merged particles. The larger mass
(merged) particle is assigned a velocity and displacement equal to the
average velocity and displacement of the smaller-mass particles.
Extensive tests of the code and comparisons with other numerical
$N$-body codes can be found in \cite{kravtsov99,knebe_etal00}.

With increasing number of particles and resolution (i.e., number of
refinement cells) the memory as well as the computing time requirement
of the code increase. In practice, on nodes with 16 Gb of shared
memory and up to 16 CPUs we are limited to simulations with $\leq
256^3$ particles 
(runtime is CPU-bound). 
The only way to overcome this
problem is to use MPI to distribute computations accross nodes.  We
have developed two different MPI algorithms to handle larger
simulations.  In the first approach, we select spatially distinct
objects of interest and simulate each of these objects on one node
with high mass and force resolution whereas in the remaining part of
the simulation box lower resolutions will be used. Each node uses then
standard ART with OpenMP on shared memory. The inter-node
communication is minimal.  In the second approach, one can divide the
whole simulation box into a number of sub-boxes which will be handled
by different nodes. Each node simulates its own sub-box with high mass
and force resolution whereas other sub-boxes have lower mass and force
resolution. Again one integration step is done by standard ART with
OpenMP on each node.  After each integration step nodes exchange
communications on positions, velocities and masses of particles that
cross sub-box boundaries.

\smallskip

\subsection{Constrained simulations of the local universe}
\label{sec:constrained}

Here our goal is to perform simulations that match the observed
local universe as well as possible. Namely, we are interested in
reproducing the observed structures: the Virgo cluster, the Local
Supercluster and the Local Group, in the approximately correct
locations and embedded within the observed large-scale configuration
dominated by the Great Attractor and Perseus-Pisces superclusters.  

An efficient algorithm for reconstructing the density and velocity
fields from sparse and noisy data of redshift and velocity surveys is
provided by the Wiener filter formalism and constrained realizations
of gaussian random fields \cite{zaroubi_etal95}. Here we use the MARK
III catalog \cite{willick_etal97}. The sample consists of $\approx
3400$ galaxies and provides radial velocities and inferred distances
with fractional errors $\sim 17-21\%$. A detailed analysis of the
large scale structure in the Local Universe reconstructed from the
MARK III survey was presented in \cite{zaroubi_etal99}.

Several simulations with increasing force and mass resolution in the
region around the Virgo Cluster were performed \cite{CR01,kkh02}.  The
initial conditions for these simulations were set using multiple mass
resolution technique. Using $z=0$ output of a low-resolution run,
we selected all particles within a sphere of $25\hMpc$ radius centered
on the Virgo cluster.  The mass resolution in the Lagrangian region
occupied by the selected particles was increased and additional
small-scale waves from the initial {$\Lambda$CDM} power spectrum of
perturbations were added appropriately \cite{kkbp01}. For the two
high-resolution simulations, the particle mass in the Local
Supercluster region is 8 and 64 times smaller than in the
low-resolution simulation.  The highest resolution simulation has a
particle mass of $2.5\times 10^9\hMsun$ and the maximum formal force
resolution was $2.4\hkpc$ in the Local Supercluster region.  The
results of both high-resolution simulations agree well with each other
at all resolved scales.

All major structures (the Local Supercluster, Great Attractor,
Perseus-Pisces supercluster, and Coma cluster) observed within
$100\hMpc$ around the Milky Way exist in the simulations. The positions
and morphology of these structures is, of course, fairly well
dictated by the constraints imposed on the initial conditions.  The
Local Supercluster is an elongated structure which extends over $\sim
40\hMpc$ along the SGX axis. There is a low-density ``bridge'' (of
overdensity just above the average density), which connects the Local
Supercluster with the Perseus-Pisces Super-cluster. There is also an
even weaker filament connecting the Local Super-cluster with the Great
Attractor.

\begin{figure}
\centering
\includegraphics[height=8cm]{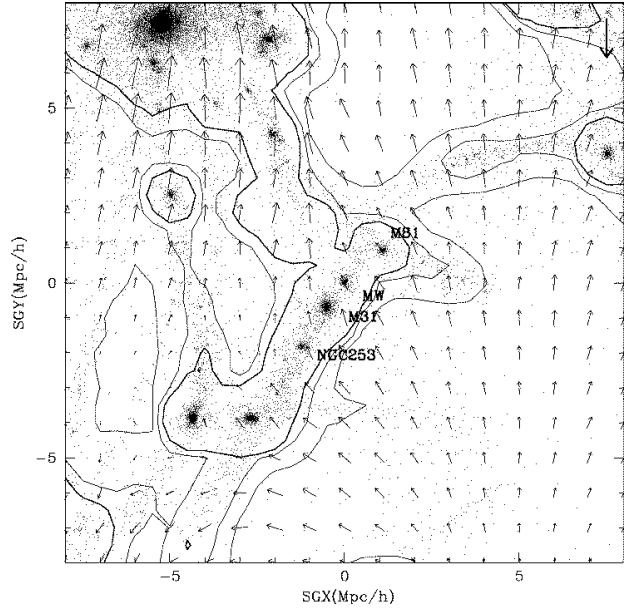}
\caption{Density (contours corresponding to overdensities of 1, 2, and 3) and velocity (arrows) fields smoothed with
  a Gaussian filter of $0.7\hMpc$ smoothing length around the Local
  Group.  The slice shown has a size and thickness of $15\hMpc$ and
  $5\hMpc$, respectively, and is centered on the supergalactic plane
  ($SGZ=0$). 
  The length of the thick arrow in the top right corner corresponds to
  a velocity of 500~km/s. The velocities are plotted in the Virgo
  cluster rest frame. }
\label{h009za_CR}       
\end{figure}

Just as in the real Universe, the Local Group is located in a weak
filament extending between the Virgo and Fornax clusters. This
filament is a counterpart of the Coma-Sculptor ``cloud'' in the
distribution of nearby galaxies.
Figure~\ref{h009za_CR} shows a zoom-in view of the immediate
environment of the simulated Local Group. Note that the structures at
these scales are only weakly affected by constraints imposed on the
initial conditions. Several possible counterparts to existing objects
(e.g., the MW and M31, M51, NGC253) are marked, but their existence is
largely fortuitous. As can be seen in this figure, the simulated Local
Group is located in a rather weak filament extending to the Virgo
cluster. This filament borders an underdense region visible in the
right lower corner of Figure~\ref{h009za_CR}, which corresponds to the
Local Void in the observed distribution of nearby galaxies. We are now
carrying out a series of high-resolution simulations of such voids
(see below). The velocity field around the Local Group is rather
quiet, in good agreement with observations.  The peculiar velocity
field in the Local Void exhibits a uniform expansion of matter out of
this underdense region, while velocities between the Local Group and
the Virgo (upper half of Fig.~\ref{h009za_CR}) show a coherent flow
onto the main body of the Local Supercluster.

\smallskip
\subsection{Galaxy clusters}

\begin{figure}
\centering
\includegraphics[width=11cm]{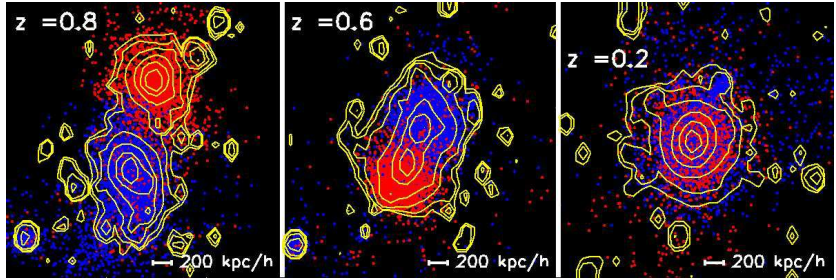}
\caption{Merger event between two $6\times10^{13}$ M$_\odot$ clusters. 
Particles belonging to one of the two clusters before merging are
plotted with red and blue dots, respectively. Yellow lines correspond
to density contours.}
\label{h009za_cluster}       
\end{figure}

Within a low mass resolution simulation ($128^3$ particles, $m_{part} =
2.0 \times 10^{10}\hMsun$) we have identified 15 clusters of galaxies
with masses above $1.0\times 10^{14}\hMsun$. From this set we have
selected 8 candidates with different masses and merging histories and
added 5 smaller clusters/groups for load balance. We
then re-simulated the clusters with higher mass resolution. With particle
masses of $3.2 \times 10^{8}\hMsun$ a typical cluster and its
environment contains more than 1 million particles. The 
formal force resolution with 9 refinement levels was $1 \hkpc$. Halos with masses
above $3.0 \times 10^{10}\hMsun$ are well resolved. A typical cluster
contains more than 150 such halos. The simulations were done using an
MPI version of the ART code where each of 8 nodes followed the
evolution of one or two clusters. We are now carrying out simulations
of these clusters including gasdynamics. 

The dynamical evolution of a typical major merger between galaxy
clusters can be seen in Fig.~\ref{h009za_cluster}. The figure shows
three snapshots of a cluster taken at $z=0.8$, $z=0.6$, and $z=0.2$.
Although the cluster does not show significant substructure in its
density contours during the merger, it is far from a relaxed state.
Most of the energy transfer takes place during the first encounter
(between $z=0.8$ and $z=0.6 $), but the merger remnant does not reach
virial equilibrium until it loses all the information about the
initial conditions (i.e. the particles are well mixed). For this
event, virialisation occurs at $z\sim 0.2$ ($\sim4$ Gyr after the
first encounter). During the first stages of cluster formation, the
characteristic time between major mergers can be shorter than the
relaxation time. High-redshift clusters may thus be in general far
from the virial equilibrium inside their formal virial radius.

\smallskip
\subsection{Voids}
\label{h009za_voids}

\begin{figure}
\centering
\vspace{-0.2cm}
\includegraphics[width=9cm]{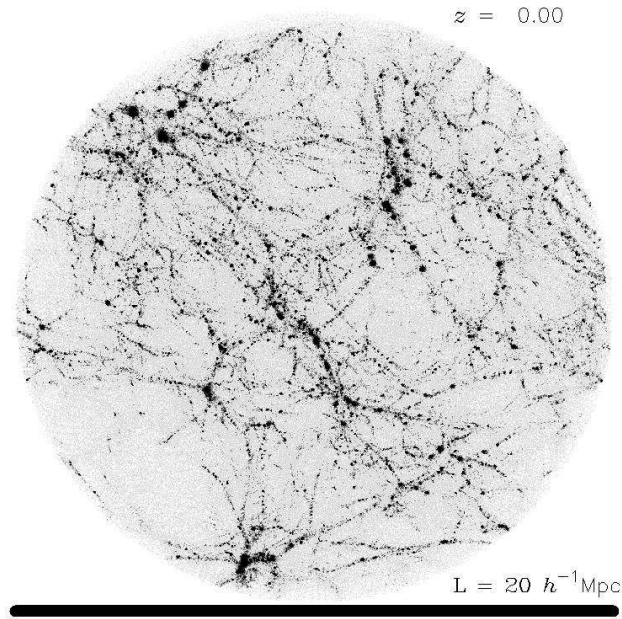}
\vspace{-0.3cm}
\caption{The inner region of a  spherical void at $z=0$.}
\label{h009za_void}       
\vspace{-0.4cm}
\end{figure}
Cosmological simulations predict many more small DM halos than the
observed number of satellites around the Milky Way and Andromeda
galaxies.  Do we have the same problem for dwarf galaxies in voids?
One naively expects a large number because the Press-Schechter mass
function steeply rises with declining mass. In contrast, it seems that
observations are failing to find a substantial number of dwarf
galaxies inside voids (e.g. \cite{pope97, GG00}).  However, the
situation is complicated because it is very difficult to detect dwarf
galaxies, many of them are expected to have low surface brightness.

In order to study the formation of large voids, the simulation box
should be sufficiently large; we use a cube of size 80\hMpc. On
the other hand, we are interested in the formation of small structure
elements inside voids, for which we need highest possible mass
resolution. Therefore, we use particles with large masses to follow the
evolution of large scale structures and particles with small masses to
follow the evolution of structure within one spherical void.

Within a low mass resolution simulation ($128^3$ particles, $m_{part} =
2.0 \times 10^{10}\hMsun$) we have identified 8387 galactic halos with
masses $>2.0 \times 10^{11} \hMsun$.  This corresponds to a mean
distance of about $4 \hMpc$ between halos. We then searched voids in
the distribution of these galactic halos by constructing
the minimal spanning tree using halo positions. The minimal
spanning tree was then used to search for the point in the simulation box
which has the largest distance $r_1$ to the set of halos. We identified
this point as the center of spherical void with the radius of $r_1$. 
Excluding that
void we were searching again for the point with the largest distance to
the set and thus found the second largest void and so on. The algorithm
is similar to that used by \cite{einasto:voids}. 

With the algorithm described above we find spherical voids in the halo
distribution which do not contain any halo with a mass larger than
$2.0 \times 10^{11} \hMsun$; such halos by definition lie on the
border of the void. After finding voids in the low-resolution
simulation, we re-run the simulation with much higher mass resolution
inside the voids ($m_{\rm part} = 4.0 \times 10^7 \hMsun$), which allows
identification of objects with masses larger than $10^9 \hMsun$. 
By construction, the voids do not have halos with masses larger
than $M_{\rm b} = 2.0 \times 10^{11} \hMsun$.  Five voids have been
identified and resimulated, their radii are $r_{void} = 11.6 $, 10.8,
9.4, 9.1, 9.1 {\hMpc}. Inside the voids the matter density is typically
a factor of 10 smaller than the mean density, but void properties
exhibit large differences. Some voids are very isolated: the density
within a sphere with radius 30 \hMpc\ centered on one of the void centers
reaches only 2/3 of the mean density. On the other hand, the most
prominent structure of the simulation, a galaxy cluster of $2 \times
10^{15} \hMpc$, is bordering one of the other voids. The density
outside of this void rapidly increases. The mass function in voids is
about an order of magnitude lower and its shape is different than that
of the field galaxies \cite{glk}.

In Fig. \ref{h009za_void} we show the inner 20 \hMpc\ of a spherical
void of radius 21.6 {\hMpc}.  In this void we found more than 50 halos
with circular velocity $v_c>50$ km/s and more than 600 halos with
$v_c>20$ km/s. There is a certain spatial mass segregation among
halos.  Typically, more massive halos tend to be situated in the outer
part of the void. The largest halos in the plot are actually
dwarf-size halos with circular velocities of $\sim$50km/s. The void is far 
from being empty and boring: it 
has a complex structure with numerous long filaments and small
sub-voids. Visually it resembles the large-scale structure of the
Universe, but everything in this plot is hundreds and thousands times
less massive than in "normal" configurations of interconnected superclusters
and filaments.

\smallskip
\subsection{Hydro simulations}

We have started to repeat \cite{kkh02} many of the simulations described above
including eulerian shock-capturing gasdynamics and physical processes
such as radiative cooling and stellar feedback. Inclusion of
gasdynamics, although significantly increasing computational and
memory demands of simulations, will allow us to address a much wider
range of questions and more robust comparison with
observations. For example, simulations of the Local Supercluster
regions with cooling and starformation will allow us to study
distribution and properties of galaxies in the simulations as a
function of their luminosity and color. High-resolution simulations of
the local voids will allow studies of spatial distribution of Ly
$\alpha$ absorbers and their connection to galaxies. Simulations of
nearby galaxy clusters (e.g., Virgo, Fornax, Coma) formed in
realistics large-scale environments will allow for detailed
object-to-object comparison of properties and should give us good
insights into physical processes operating within intracluster medium.

\begin{figure}
\centering
\includegraphics[width=4.8cm,angle=-90]{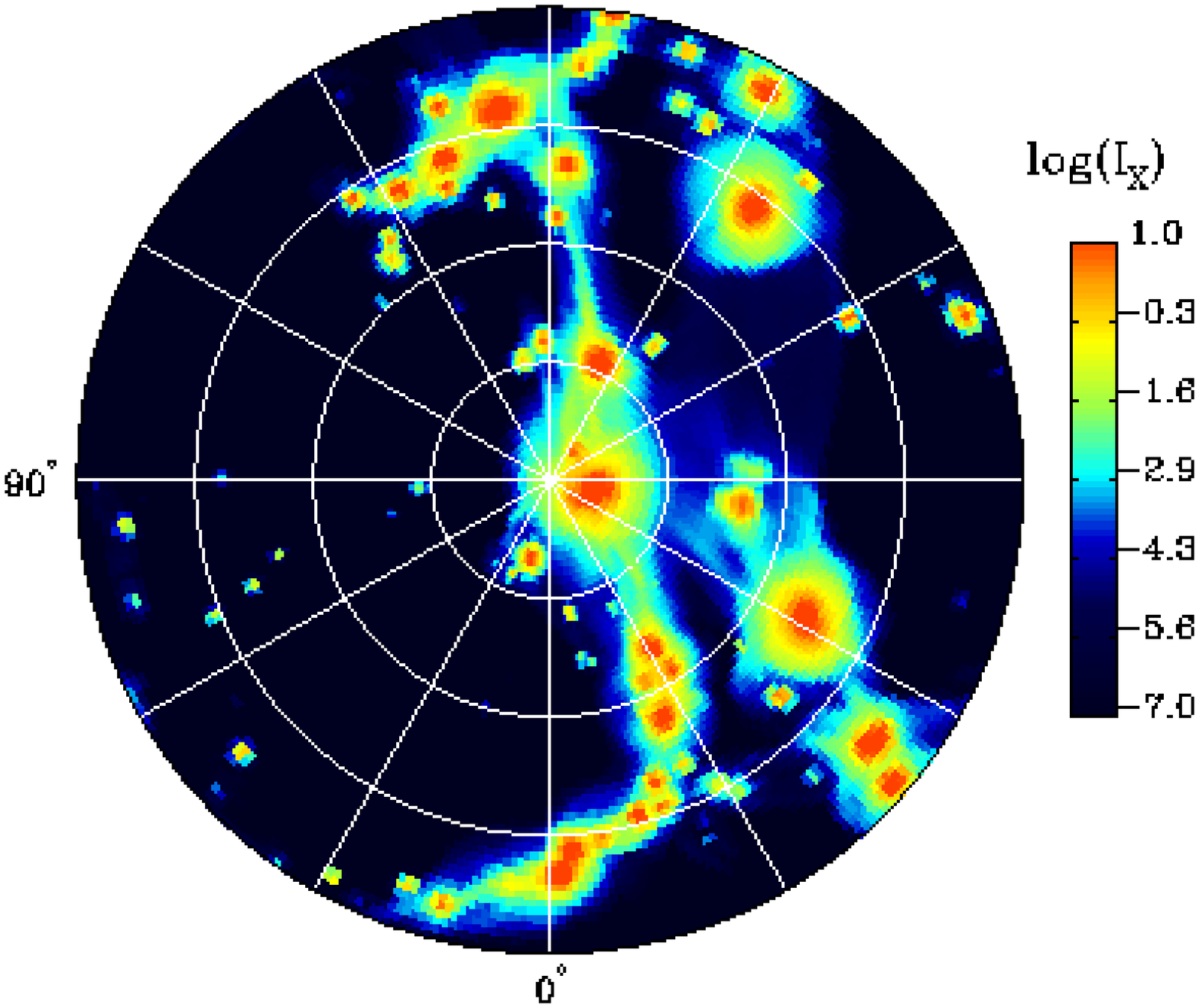}
\includegraphics[width=4.8cm,angle=-90]{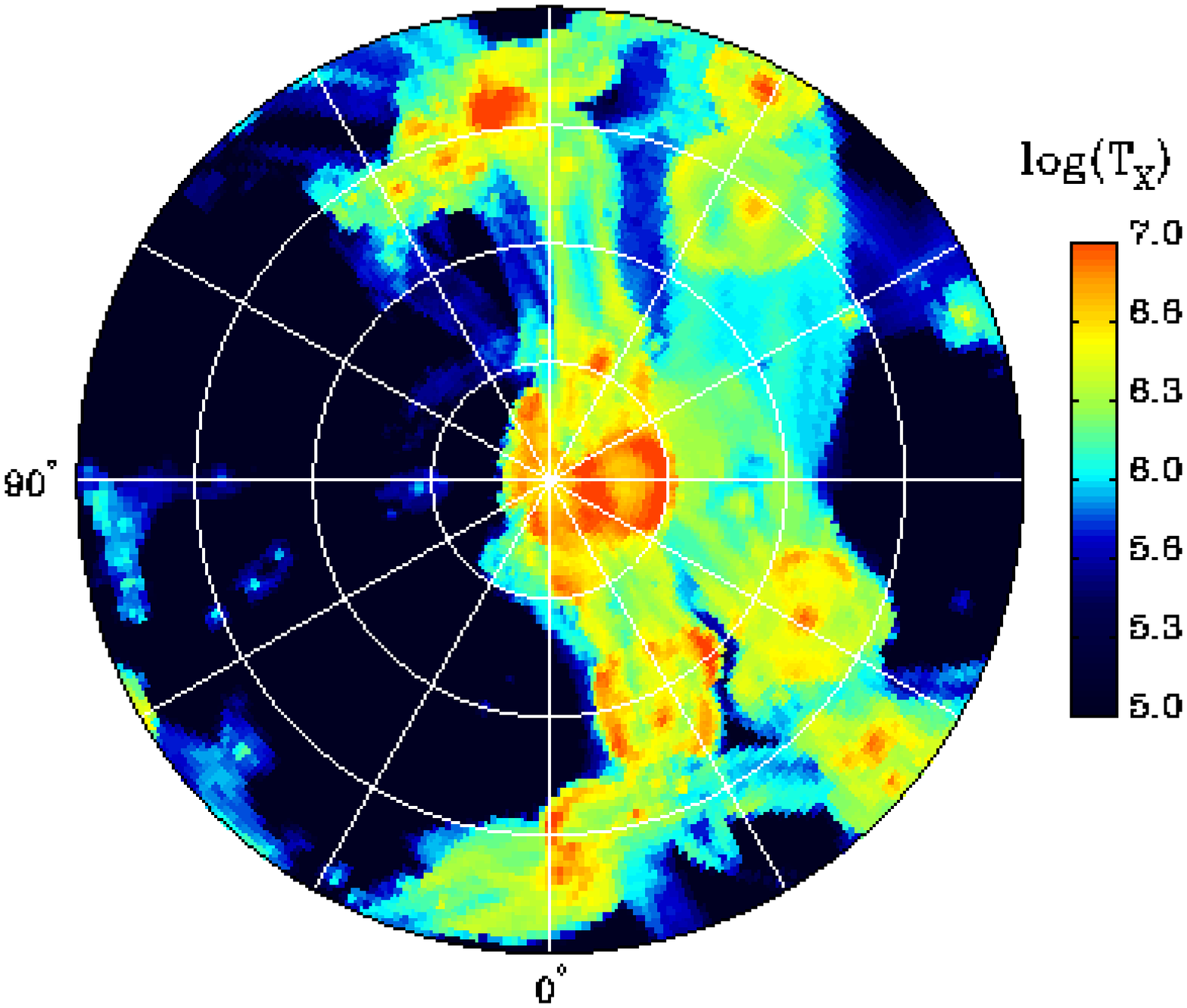}
\caption{The projected X-ray intensity and emission-weighted temperature
maps of the Local supercluster in galactic coordinates. The Virgo cluster
is the brightest object near the center (which corresponds to the North
Galactic Pole).}
\label{h009za_tlx}       
\end{figure}

\begin{figure*}
\centering
\includegraphics[width=10cm]{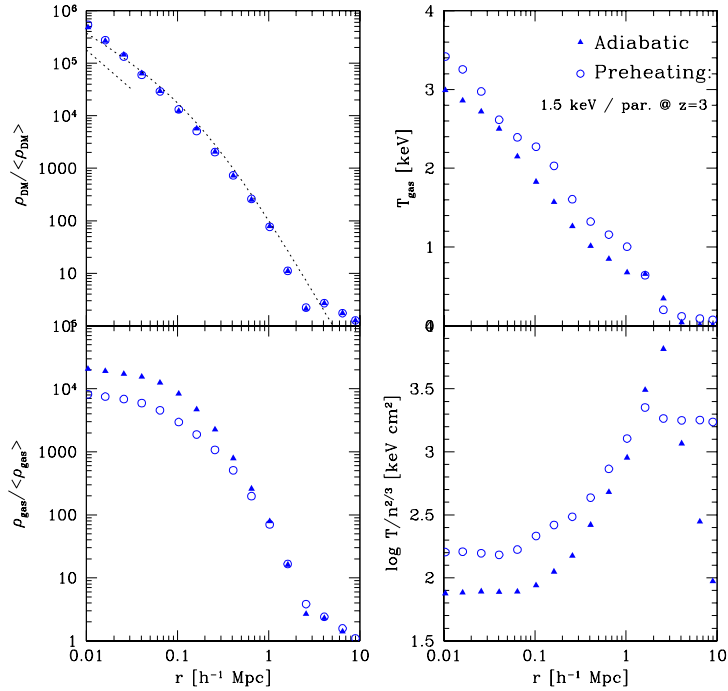}
\caption{Radial profiles of dark matter and gas density, gas temperature, 
and gas entropy of the simulated Virgo cluster. The dotted line in the
DM density panel shows the NFW profile with the concentration of 7.}
\label{h009za_vpro}       
\end{figure*}

\begin{figure}
\centering
\includegraphics[width=8.5cm]{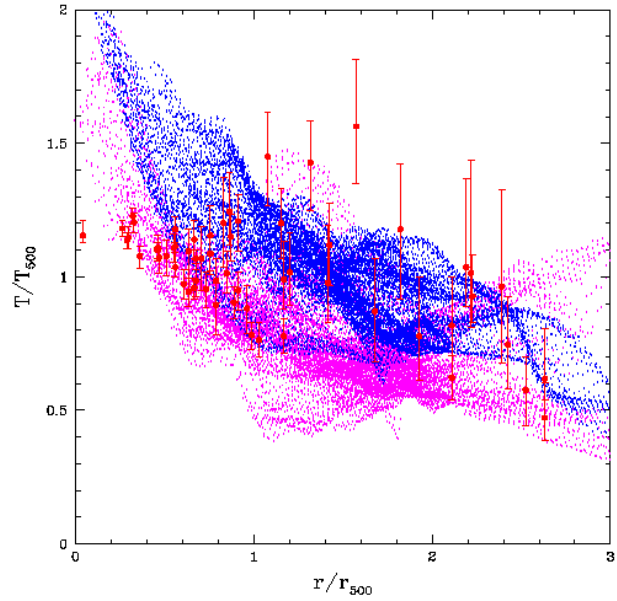}
\caption{Projected temperature profiles of the observed (points with
errorbars) and simulated (purple and blue points) Virgo cluster.
Purple points correspond to the adiabatic simulation while blue points
show temperature in the simulation with preheating. The radii are normalized
to the radius corresponding to the overdensity of 500 with respect to the critical
density.}
\label{h009za_vtpro}       
\end{figure}

As an example, we present here simulations of the Virgo cluster in the
context of constrained simulations described in
\S~\ref{sec:constrained}.  The lagrangian region within five virial
radii (at $z=0$) around the cluster was resimulated with mass
resolution 8 times higher (particle mass of $3.1\times 10^8 \hMsun$)
than that of the Local Supercluster simulation and spatial resolution
of $\approx 1\hkpc$ in the central regions of the cluster.
Figure~\ref{h009za_tlx} shows the sky projection of the X-ray
intensity and emission-weighted temperature of gas in the simulations
in galactic coordinates (the center of the polar projection
corresponds to the North Galactic Pole, NGP). The Virgo cluster is the
brightest object located near the NGP, very close to the actual
location of the real Virgo cluster. Figure~\ref{h009za_vpro} shows
radial profiles of dark matter and gas density, as well as profiles of
gas temperature and entropy for two simulations: one that included
only adiabatic gasdynamics, and the other in which gas was preheated
with the energy of 1.5 keV per gas particle at $z=3$. The preheating
very roughly models the possible effect of galactic winds at
high-redshifts.  The figure shows that preheating increases entropy of
the gas thereby lowering its density in the cluster core and
increasing the overall gas temperature.  It lowers the X-ray
luminosity of the cluster by a factor of 8, which brings it in good
agreement with the observed luminosity-temperature relation.  Note,
however, that it does not change the shape of the temperature profile.
The proximity of the Virgo cluster makes it one of the best spatially
resolved clusters, with the temperature mesurements well outside the
virial radius.  Figure~\ref{h009za_vtpro} compares the observed
temperature measurements around the center of the Virgo cluster
\cite{shibata01} as a function of projected radius to the
corresponding measurements in simulations (adiabatic and preheating).
We constructed projected map of emission-weighted temperature of the
simulated Virgo cluster using $4^{\prime}\times 4^{\prime}$ pixels,
each pixel represented by a point in the figure.  The figure shows
that neither simulation can match the much shallower temperature
distribution in the central regions of the Virgo cluster. This
indicates that preheating alone cannot be the whole story and other
processes, such as cooling and/or central heating by AGNs can be
important. We will be exploring the effects of these processes in
future simulations.

\smallskip 
\section{Summary}

The adaptive refinement tree code is a useful tool to study
cosmological structure formation on different scales and with different
resolutions. It runs well on a variety of platforms with shared and
distributed memory. The simulations described here have
been performed on the Hitachi of LRZ Munich, the small development Hitachi at
the AIP, the IBM SP of the Potsdam Institute for Climate Impact Research, and
the IBM SP of NERSC Berkeley.

%


\printindex
\end{document}